\documentclass[11pt]{article}
\usepackage{epsfig}
\usepackage{amsmath,amsfonts}

\newcommand{\MSbar}{\overline{\mbox{MS}}}

\begin{document}

\title{\textbf{Renormalization properties of the mass operator }$A_{\mu
}^{a}A_{\mu }^{a}$\textbf{\ in three dimensional Yang-Mills theories
in the Landau gauge}}
\author{\textbf{D. Dudal}$^{a}$\thanks{%
Research Assistant of The Fund For Scientific Research-Flanders,
Belgium.} \
, \textbf{J.A. Gracey}$^{b}$\thanks{%
jag@amtp.liv.ac.uk } \ , \textbf{V.E.R. Lemes}$^{c}$\thanks{%
vitor@dft.if.uerj.br} \ , \textbf{R.F. Sobreiro}$^{c}$\thanks{%
sobreiro@uerj.br}  \and  \textbf{S.P. Sorella}$^{c}$\thanks{%
sorella@uerj.br} \ , \textbf{H. Verschelde}$^{a}$\thanks{%
david.dudal@ugent.be, henri.verschelde@ugent.be} \\\\
\textit{$^{a}$ Ghent University} \\
\textit{Department of Mathematical Physics and Astronomy} \\
\textit{Krijslaan 281-S9, B-9000 Gent, Belgium}\\
[3mm] \textit{$^{b}$ Theoretical Physics Division} \\
\textit{Department of Mathematical Sciences} \\
\textit{University of Liverpool} \\
\textit{P.O. Box 147, Liverpool, L69 3BX, United Kingdom} \\
[3mm] \textit{$^{c}$ UERJ, Universidade do Estado do Rio de Janeiro} \\
\textit{Rua S{\~a}o Francisco Xavier 524, 20550-013 Maracan{\~a}} \\
\textit{Rio de Janeiro, Brasil}} \maketitle

\begin{abstract}
Massive renormalizable Yang-Mills theories in three dimensions are
analysed within the algebraic renormalization in the Landau gauge.
In analogy with the four dimensional case, the renormalization of
the mass operator $A_{\mu }^{a}A_{\mu }^{a}$ turns out to be
expressed in terms of the fields and coupling constant
renormalization factors. We verify the relation we obtain
for the operator anomalous dimension by explicit calculations in the large $%
N_{\!f}$ expansion. The generalization to other gauges such as the
nonlinear Curci-Ferrari gauge is briefly outlined.
\end{abstract}

\vspace{-18cm} \hfill LTH--630 \vspace{18cm}

\newpage

\section{Introduction.}
Recently, much work has been devoted to the study of the operator
$A_{\mu }^{a}A_{\mu }^{a}$ in four dimensional Yang-Mills theories
in the Landau gauge, where a renormalizable effective potential for
this operator can be consistently constructed
\cite{Verschelde:2001ia,Browne:2003uv}. This has produced analytic
evidence of a nonvanishing condensate $\left\langle A_{\mu
}^{a}A_{\mu }^{a}\right\rangle $, resulting in a dynamical mass
generation for the gluons \cite{Verschelde:2001ia,Browne:2003uv}. A
gluon mass in the Landau gauge has been reported in lattice
simulations \cite{Langfeld:2001cz} as well as in a recent
investigation of the Schwinger-Dyson equations
\cite{Aguilar:2004kt}. Besides being multiplicatively renormalizable
to all orders of perturbation theory in the Landau gauge, the
operator $A_{\mu }^{a}A_{\mu }^{a}$ displays remarkable properties.
In fact, it has been proven \cite{Dudal:2002pq} by using BRST\ Ward
identities that the anomalous dimension $\gamma _{A^2}(a)$ of the
operator $A_{\mu }^{a}A_{\mu }^{a}$ in the Landau gauge is not an
independent parameter, being expressed as a combination of the gauge
beta function, $\beta (a)$, and of the anomalous dimension, $\gamma
_{A}(a)$, of the gauge field, according to the relation
\begin{equation}
\gamma _{A^{2}}(a)~=~-~\left( \frac{\beta (a)}{a}~+~\gamma
_{A}(a)\right) ,\,\,\,\,\,\,\,\,\,\,\,a=\frac{g^{2}}{16\pi
^{2}}\,\,\,, \label{g1}
\end{equation}
which can be explicitly verified by means of the three loop
computations available in \cite{Gracey:2002yt}. The operator $A_{\mu
}^{a}A_{\mu }^{a}$ turns out to be multiplicatively renormalizable
also in the linear covariant gauges \cite{Dudal:2003np}. Its
condensation and the ensuing dynamical gluon mass generation in this
gauge have been discussed in \cite{Dudal:2003by}.

Moreover, the operator $A_{\mu }^{a}A_{\mu }^{a}$ in the Landau
gauge can be generalized to other gauges such as the Curci-Ferrari
and maximal Abelian gauges. Indeed, as was shown in
\cite{Kondo:2001nq,Kondo:2001tm}, the mixed
gluon-ghost operator\footnote{%
In the case of the maximal Abelian gauge, the color index $a$ runs
only over the $N(N-1)$ off-diagonal components.} $\left(
\frac{1}{2}A_{\mu }^{a}A_{\mu }^{a}+\alpha
\overline{c}^{a}c^{a}\right) $ turns out to be BRST\ invariant
on-shell, where $\alpha $ is the gauge parameter. In both gauges,
the
operator $\left( \frac{1}{2}A_{\mu }^{a}A_{\mu }^{a}+\alpha \overline{c}%
^{a}c^{a}\right) $ turns out to be multiplicatively renormalizable
to all orders of perturbation theory and, as in the case of the
Landau gauge, its anomalous dimension is not an independent
parameter of the theory \cite {Dudal:2003pe}. A detailed study of
the analytic evaluation of the effective potential for the
condensate $\left\langle \frac{1}{2}A_{\mu }^{a}A_{\mu }^{a}+\alpha
\overline{c}^{a}c^{a}\right\rangle $ in these gauges can be found in
\cite{Dudal:2003gu,Dudal:2004rx}. In particular, it is worth
emphasizing that in the case of the maximal Abelian gauge, the
off-diagonal
gluons become massive due to the gauge condensate $\left\langle \frac{1}{2}%
A_{\mu }^{a}A_{\mu }^{a}+\alpha \overline{c}^{a}c^{a}\right\rangle
$, a fact that can be interpreted as evidence for the Abelian
dominance hypothesis underlying the dual superconductivity mechanism
for color confinement.

The aim of this work is to analyse the renormalization properties of
the operator $A_{\mu}^{a}A_{\mu }^{a}$ in three dimensional
Yang-Mills theories in the Landau gauge. This investigation might be
useful in order to study by analytical methods the formation of the
condensate $\left\langle A_{\mu }^{a}A_{\mu }^{a}\right\rangle $ in
three dimensions, whose relevance for the Yang-Mills theories at
high temperatures has been pointed out long ago
\cite{Appelquist:1981vg}. Furthermore, the possibility of a
dynamical gluon mass generation related to the operator $A_{\mu
}^{a}A_{\mu }^{a}$ could provide a suitable infrared cutoff which
would prevent three dimensional Yang-Mills theory from the well
known infrared instabilities \cite {Jackiw:1980kv}, due to its
superrenormalizability.

The organization of the paper is as follows. In Sect.2 we discuss
the renormalizability of the three dimensional Yang-Mills  theory
in the Landau gauge, when the operator $A_{\mu }^{a}A_{\mu
}^{a}\;$is added to the starting action in the form of a mass term,
$m^{2}\int d^{3}xA_{\mu }^{a}A_{\mu }^{a}$. We shall be able to
prove that the renormalization factor $Z_{m^{2}}$ of the mass
parameter $m^{2}$ can be expressed in terms of the renormalization
factors $Z_{A}$ and $Z_{g}$ of the gluon field and of the gauge
coupling constant, according to
\begin{equation}
Z_{m^{2}}=Z_{g}Z_{A}^{-1/2}\;.  \label{zmm}
\end{equation}
This relation represents the analogue in three dimensions of the
eq.$\left( \ref{g1}\right) $. In Sect.3 we give an explicit
verification of the relation $\left( \ref{zmm}\right) $ by using the
large $N_{\!f}$ expansion method. In Sect.4 we present the
generalization to the nonlinear Curci-Ferrari gauge.

\section{Renormalizability of massive three dimensional Yang-Mills theory in
the Landau gauge.}
\subsection{Ward identities.}
In order to analyze the renormalizability of three dimensional
Yang-Mills theory, in the presence of the mass term
$\frac{1}{2}m^{2}\int d^{3}xA_{\mu }^{a}A_{\mu }^{a}$, we start from
the following gauge fixed action
\begin{equation}
S=\int d^{3}x\,\left( -~\frac{1}{4}F_{\mu \nu }^{a}F_{\mu \nu }^{a}+\frac{1}{%
2}m^{2}A_{\mu }^{a}A_{\mu }^{a}+b^{a}\partial _{\mu }A_{\mu }^{a}+\,%
\overline{c}^{a}\partial _{\mu }\left( D_{\mu }c\right) ^{a}\right)
\;, \label{si}
\end{equation}
with
\begin{equation}
\left( D_{\mu }c\right) ^{a}~=~\partial _{\mu
}c^{a}~+~gf^{abc}A_{\mu }^{b}c^{c}\;,  \label{cder}
\end{equation}
where $b^{a}$ is the Lagrange multiplier enforcing the Landau gauge
condition, $\partial _{\mu }A_{\mu }^{a}=0$, and $\overline{c}^{a}$,
$c^{a}$ are the Faddeev-Popov ghosts. Concerning the mass term in
expression $\left( \ref{si}\right) $, two remarks are in order. The
first one is that, although in three dimensions the gauge field
might become massive due to the introduction of the Chern-Simons
topological action \cite{Deser:1981wh}, one should note that the
mass term considered here is of a different nature. In fact, unlike
the Chern-Simons term, the mass term $m^{2}A_{\mu }^{a}A_{\mu }^{a}$
does not break parity. As a consequence, the starting action $\left(
\ref{si}\right) $ is parity preserving. Therefore, the parity
breaking Chern-Simons term cannot show up due to radiative
corrections. The second remark is related to the
superrenormalizabilty of three dimensional Yang-Mills theories, as
expressed by the dimensionality of the gauge coupling $g$. As shown
in \cite{Jackiw:1980kv}, a standard perturbation theory would be
affected by infrared singularities in the massless case. However,
the presence of the mass term prevents the theory from this infrared
instability, allowing one to define an infrared safe perturbative
expansion.

Following \cite{Delduc:1989uc}, the action $\left( \ref{si}\right)$
is left invariant by a set of modified BRST\ transformations, given
by
\begin{eqnarray}
&&sA_{\mu }^{a}~=-\left( D_{\mu }c\right) ^{a}\,\,,\,\,\;\;sc^{a}~=\frac{g}{2%
}f^{abc}c^{b}c^{c}\;,  \nonumber \\
&&s\overline{c}^{a}~=b^{a}\,\,,\hspace{1.6cm}sb^{a}~=-m^{2}c^{a}\;,
\label{sm}
\end{eqnarray}
and
\begin{equation}
sS=0\;.  \label{sinv}
\end{equation}
Notice that, due to the introduction of the mass term $m$, the
operator $s$ is not strictly nilpotent, \textit{i.e. }
\begin{eqnarray}
&&s^{2}\Phi ~=~0\;,\,\,\,\;(\Phi =A^{a\mu },\,c^{a})\;,  \nonumber \\
&&s^{2}\overline{c}^{a}~=-~m^{2}c^{a}\,,  \nonumber \\
&&s^{2}b^{a}~=-m^{2}\frac{g}{2}f^{abc}c^{b}c^{c}\,.  \label{s2}
\end{eqnarray}
Therefore, setting
\begin{equation}
s^{2}~\equiv -m^{2}~\delta \;,  \label{d}
\end{equation}
we have
\begin{equation}
\delta S=0\;.  \label{di}
\end{equation}
The operator $\delta $ is related to a global $SL(2,\mathbb{R})$
symmetry \cite{Delduc:1989uc}, which is known to be present in the
Landau, Curci-Ferrari and maximal Abelian gauges
\cite{Dudal:2002ye}. Finally, in order to express the BRST\ and
$\delta $ invariances in a functional way, we introduce the external
action \cite{Piguet:1995er}
\begin{eqnarray}
S_{\mathrm{ext}} &=&\int d^{3}x\left( \Omega _{\mu }^{a}\,sA_{\mu
}^{a}\,+\,L^{a}\,sc^{a}\right)  \label{sext} \\
&=&\int d^{3}x\left( -~\Omega _{\mu }^{a}\,\left( D_{\mu }c\right)
^{a}+\,L^{a}\,\frac{g}{2}f^{abc}c^{b}c^{c}\right) ,  \nonumber
\end{eqnarray}
where $\Omega _{\mu }^{a}$ and $L^{a}$ are external sources
invariant under both BRST and $\delta$ transformations, coupled to
the nonlinear variations of the fields $A_{\mu }^{a}$ and $c^{a}$.
It is easy to check that the complete classical action,
\begin{equation}
\Sigma ~=~S+~S_{\mathrm{ext}}\;,  \label{ca}
\end{equation}
is invariant under BRST and $\delta $ transformations
\begin{equation}
s\Sigma ~=~0\,\,\,,\,\,\,\,\delta \Sigma ~=~0\;.  \label{inv}
\end{equation}
When translated into functional form, the BRST and the $\delta $
invariances
give rise to the following Ward identities for the complete action $\Sigma $%
, namely
\begin{itemize}
\item  the Slavnov-Taylor identity
\end{itemize}
\begin{equation}
\mathcal{S}(\Sigma ) ~=~ 0\;,  \label{sti}
\end{equation}
\begin{itemize}
\item  with
\begin{equation}
\mathcal{S}(\Sigma )=\int d^{3}x\,\left( \frac{\delta \Sigma
}{\delta \Omega
_{\mu }^{a}}\frac{\delta \Sigma }{\delta A_{\mu }^{a}}+\frac{\delta \Sigma }{%
\delta L^{a}}\frac{\delta \Sigma }{\delta c^{a}}+b^{a}\frac{\delta \Sigma }{%
\delta \overline{c}^{a}}-m^{2}c^{a}\frac{\delta \Sigma }{\delta b^{a}}%
\right) \;,  \label{std}
\end{equation}
\item  the $\delta $ Ward identity
\begin{equation}
\mathcal{W}\left( \Sigma \right) ~=~0\;,  \label{wwi}
\end{equation}
with
\begin{equation}
\mathcal{W}(\Sigma )=\int d^{3}x\,\left( c^{a}\frac{\delta \Sigma
}{\delta
\overline{c}^{a}}+\frac{\delta \Sigma }{\delta L^{a}}\frac{\delta \Sigma }{%
\delta b^{a}}\right) \;.  \label{wwd}
\end{equation}
\end{itemize}
In addition, the following Ward identities holds in the Landau gauge
\cite{Piguet:1995er}, \textit{i.e.}
\begin{itemize}
\item  the gauge fixing condition and the antighost equation
\begin{equation}
\frac{\delta \Sigma }{\delta b^{a}}=\partial _{\mu }A_{\mu
}^{a}\;,\;\;\;\;\;\;\;\;\frac{\delta \Sigma }{\delta \overline{c}^{a}}%
+\partial _{\mu }\frac{\delta \Sigma }{\delta \Omega _{\mu
}^{a}}=0\;, \label{beq}
\end{equation}
\item  the integrated ghost equation \cite{Blasi:1990xz,Piguet:1995er}
\begin{equation}
\mathcal{G}^{a}\Sigma =\Delta _{\mathrm{cl}}^{a}\;,  \label{gh}
\end{equation}
with
\begin{equation}
\mathcal{G}^{a}=\int d^{3}x\,\left( \frac{\delta }{\delta c^{a}}+gf^{abc}%
\overline{c}^{b}\frac{\delta }{\delta b^{c}}\right) \;,
\label{ghop}
\end{equation}
and
\begin{equation}
\Delta _{\mathrm{cl}}^{a}\;=g\int d^{3}xf^{abc}\left( A_{\mu
}^{b}\Omega _{\mu }^{c}-L^{b}c^{c}\right) \;.  \label{br}
\end{equation}
Notice that the breaking term $\Delta _{\mathrm{cl}}^{a}$ in the
right-hand side of eq.$\left( \ref{gh}\right) $, being linear in the
quantum fields, is a classical breaking, not affected by quantum
corrections \cite{Blasi:1990xz,Piguet:1995er}.
\end{itemize}

\subsection{Algebraic characterization of the invariant counterterm.}
Having established all Ward identities obeyed by the classical action $%
\Sigma $, we can now proceed with the characterization of the most
general
local counterterm compatible with the identities (\ref{sti}), (\ref{wwi}), (%
\ref{beq}) and (\ref{gh}). Let us begin by displaying the quantum
numbers of all fields, sources and parameters
\begin{equation}
\begin{tabular}{|c|c|c|c|c|c|c|c|c|c|}
\hline
& $A_{\mu }^{a}$ & $c^{a}$ & $\overline{c}^{a}$ & $b^{a}$ & $L^{a}$ & $%
\Omega _{\mu }^{a}$ & $g$ & $s$ & $m$ \\ \hline Gh. number & $0$ &
$1$ & $-1$ & $0$ & $-2$ & $-1$ & $0$ & $1$ & $0$ \\ \hline
Dimension & $1/2$ & $0$ & $1$ & $3/2$ & $5/2$ & $2$ & $1/2$ & $1/2$ & $1$ \\
\hline
\end{tabular}
\label{t1}
\end{equation}
In order to characterize the most general invariant counterterm
which can be freely added to all orders of perturbation theory, we
perturb the classical action $\Sigma $ by adding an arbitrary
integrated, parity preserving, local polynomial $\Sigma
^{\mathrm{count}}$ in the fields and external sources of dimension
bounded by three and with zero ghost number, and we require that the
perturbed action $(\Sigma +\eta \Sigma ^{\mathrm{count}})$ satisfies
the same Ward identities and constraints as $\Sigma $ to first order
in the perturbation parameter $\eta $, which are
\begin{eqnarray}
\mathcal{S}(\Sigma +\eta \Sigma ^{\mathrm{count}})
&=&0~+~O(\eta ^{2})\;,  \nonumber \\
\mathcal{W}\left( \Sigma +\eta \Sigma ^{\mathrm{count}}\right)
&=&0~+~O(\eta ^{2})\;,  \nonumber \\
\frac{\delta \left( \Sigma +\eta \Sigma ^{\mathrm{count}}\right) }{%
\delta b^{a}} &=&\partial _{\mu }A_{\mu }^{a}\;+O(\eta ^{2})\;,
\nonumber \\
\left( \frac{\delta }{\delta \overline{c}^{a}}+\partial _{\mu
}\frac{\delta
}{\delta \Omega _{\mu }^{a}}\right) \left( \Sigma +\eta \Sigma ^{%
\mathrm{count}}\right) &=&0\;+O(\eta ^{2})\;,  \nonumber \\
\mathcal{G}^{a}\left( \Sigma +\eta \Sigma ^{\mathrm{count}}\right)
&=&\Delta _{\mathrm{cl}}^{a}\;+O(\eta ^{2})\;.\;  \label{eps}
\end{eqnarray}
This amounts to imposing the following conditions on $\Sigma ^{\mathrm{count}%
}$%
\begin{equation}
\mathcal{B}_{\Sigma }\Sigma ^{\mathrm{count}}~=~0\;,  \label{stt}
\end{equation}
with
\begin{eqnarray}
\mathcal{B}_{\Sigma } &=&\int d^{3}x\left( \frac{\delta \Sigma
}{\delta
A_{\mu }^{a}}\frac{\delta }{\delta \Omega ^{a\mu }}+\frac{\delta \Sigma }{%
\delta \Omega ^{a\mu }}\frac{\delta }{\delta A_{\mu
}^{a}}+\frac{\delta
\Sigma }{\delta L^{a}}\frac{\delta }{\delta c^{a}}+\frac{\delta \Sigma }{%
\delta c^{a}}\frac{\delta }{\delta L^{a}}\right.  \nonumber \\
&+&\left. b^{a}\frac{\delta }{\delta \overline{c}^{a}}%
-m^{2}c^{a}\frac{\delta }{\delta b^{a}}\right) \;,  \label{bs}
\end{eqnarray}
\begin{equation}
\mathcal{W}_{\Sigma }\Sigma ^{\mathrm{count}}=\int d^{3}x\,\left( c^{a}\frac{%
\delta \Sigma ^{\mathrm{count}}}{\delta
\overline{c}^{a}}+\frac{\delta
\Sigma }{\delta L^{a}}\frac{\delta \Sigma ^{\mathrm{count}}}{\delta b^{a}}+%
\frac{\delta \Sigma }{\delta b^{a}}\frac{\delta \Sigma ^{\mathrm{count}}}{%
\delta L^{a}}\right) ~=~0\;,  \label{dw}
\end{equation}
\begin{equation}
\frac{\delta \Sigma ^{\mathrm{count}}}{\delta b^{a}}=0\;,\;\;\;\;\;\;\frac{%
\delta \Sigma ^{\mathrm{count}}}{\delta \overline{c}^{a}}+\partial _{\mu }%
\frac{\delta \Sigma ^{\mathrm{count}}}{\delta \Omega _{\mu
}^{a}}=0\;, \label{bc}
\end{equation}
and
\begin{equation}
\mathcal{G}^{a}\Sigma ^{\mathrm{count}}=0\;.  \label{gc}
\end{equation}
Following the algebraic renormalization procedure
\cite{Piguet:1995er}, it turns out that the most general local,
parity preserving, invariant counterterm $\Sigma ^{\mathrm{count}}$
compatible with all constraints (\ref {stt}), (\ref{dw}), (\ref{bc})
and (\ref{gc}), contains only two independent free parameters
$\sigma $ and $a_{1}$, and is given by
\begin{eqnarray}
\Sigma ^{\mathrm{count}} &=&\int d^{3}x\,\left( -\frac{\left( \sigma
+4a_{1}\right) }{4}F_{\mu \nu }^{a}F_{\mu \nu }^{a}+a_{1}F_{\mu \nu
}^{a}\partial _{\mu }A_{\nu }^{a}+\frac{a_{1}}{2}m^{2}A_{\mu
}^{a}A_{\mu
}^{a}\right.  \nonumber \\
&+&\left. a_{1}\left( \Omega _{\mu }^{a}+\partial _{\mu }%
\overline{c}^{a}\right) \partial _{\mu }c^{a}\right) \;.  \nonumber \\
&&  \label{sc}
\end{eqnarray}

\subsection{Stability and renormalization of the mass parameter.}
It remains now to discuss the stability of the classical action
\cite{Piguet:1995er}, \textit{i.e.} to check that $\Sigma
^{\mathrm{count}}$ can be reabsorbed in the classical action $\Sigma
$ by means of a multiplicative renormalization of the coupling
constant $g$, the mass parameter $m^{2}$,
the fields $\left\{ \phi =A,c,\overline{c},b\right\} $ and the sources $%
L,\;\Omega $, namely
\begin{equation}
\Sigma (g,m^{2},\phi ,L,\Omega )+\eta \Sigma
^{\mathrm{count}}=\Sigma (g_{0},m_{0}^{2},\phi _{0},L_{0},\Omega
_{0})+O(\eta ^{2})\;, \label{stab}
\end{equation}
with the bare fields and parameters defined as
\begin{eqnarray}
A_{0\mu }^{a} &=&Z_{A}^{1/2}A_{\mu }^{a}\,\,,\,\;\;\;\;\Omega _{0\mu
}^{a}~=~Z_{\Omega }\Omega _{\mu }^{a}\,\,\,,\,\,\,  \nonumber \\
c_{0}^{a}
&=&Z_{c}^{1/2}c^{a}\,,\,\,\,\;\;\;\;\;\;\;L_{0}^{a}~=~Z_{L}L^{a}\,\,\,\,\,,%
\,\,\,  \nonumber \\
\;g_{0}~
&=&~Z_{g}g\,\,,\;\;\;\;\;\;\;\;\;\;\;\;m_{0}^{2}=Z_{m^{2}}m^{2}\;,
\nonumber \\
\overline{c}_{0}^{a} &=&Z_{\overline{c}}^{1/2}\overline{c}%
^{a}\,,\,\,\,\;\;\;\;\;\;\;\;\;b_{0}^{a}=Z_{b}^{1/2}b^{a}\,\,\,\,.\,\,\,
\label{r}
\end{eqnarray}
The parameters $\sigma $ and $a_{1}$, are easily seen to be related
to the
renormalization of the gauge coupling constant $g$ and of the gauge field $%
A_{\mu }^{a}$, according to
\begin{eqnarray}
Z_{g} &=&1-\eta \frac{\sigma }{2}\,\,,  \nonumber \\
\,Z_{A}^{1/2} &=&1+\eta \left( \frac{\sigma }{2}+a_{1}\right) .
\label{rga}
\end{eqnarray}
Concerning the other fields and the sources $\Omega _{\mu }^{a}$,
$L^{a}$, it can be verified that they are renormalized as
\begin{equation}
Z_{\overline{c}}\,=Z_{c}=Z_{g}^{-1}Z_{A}^{-1/2}\;,  \label{rgh}
\end{equation}
\begin{equation}
Z_{b}=Z_{A}^{-1}\;,\;\;\;Z_{\Omega
}~=~Z_{c}^{1/2}\;,\;\;\;Z_{L}=Z_{A}^{1/2}\;.  \label{lr}
\end{equation}
Finally, for the mass parameter $m^{2}$,
\begin{equation}
Z_{m^{2}}~=~Z_{g}\,Z_{A}^{-1/2},  \label{rm}
\end{equation}
which, due to eq.(\ref{rgh}), can be rewritten as
\begin{equation}
Z_{m^{2}}~=~Z_{c}^{-1}\,Z_{A}^{-1}\;.  \label{rm1}
\end{equation}
Equation (\ref{rgh}) expresses the well known nonrenormalization
property of the ghost-antighost-gluon vertex in the Landau gauge. As
shown in \cite{Blasi:1990xz}, this is a direct consequence of the
ghost Ward identity (\ref{gh}). Also, as anticipated, equation
(\ref{rm}) shows that the renormalization of the mass parameter
$m^{2}$ can be expressed in terms of the gauge field and coupling
constant renormalization factors. It is worth mentioning here that
eqs.(\ref{rgh}), (\ref{rm}) are in complete agreement with the
results obtained in the case of the four dimensional Yang-Mills
theory in the Landau gauge \cite{Dudal:2002pq}.

Although we did not consider matter fields in the previous analysis,
it can be easily checked that the renormalizability of the mass
operator $A_{\mu }^{a}A_{\mu }^{a}$ and the relations (\ref{rm}),
(\ref{rm1}) remain unchanged if massless spinor fields are included,
namely
\begin{equation}
S_{\mathrm{matter}}~=~\int d^{3}x\;\left( i\bar{\psi}^{i}\partial
\!\!\!/\psi ^{i}+~gA_{\mu }^{a}\bar{\psi}^{i}\gamma ^{\mu }T^{a}\psi
^{i}~\right) \;,  \label{sf}
\end{equation}
with $i=1,\ldots,N_{\!f}$. In fact, as was pointed out in
\cite{Jackiw:1980kv}, the addition of massless fermions does not
break the parity invariance of the starting action $\left(
\ref{si}\right)$. Of course, the inclusion of the matter action
(\ref{sf}) requires the introduction of a suitable renormalization
factor $Z_{\psi }$ for the spinor fields.

\subsection{Absence of one loop ultraviolet divergences.}
In the previous section we have proven that the massive three
dimensional Yang-Mills action (\ref{si}) is multiplicatively
renormalizable to all orders of perturbation theory, displaying
interesting renormalization features, as expressed by equations
(\ref{rgh}) and (\ref{rm}). Only two renormalization constants,
$Z_{g}$\ and $Z_{A}$, are needed at the quantum level. These factors
should be computed order by order by means of a suitable
regularization, which in the present case could be provided by
dimensional regularization. Due to the absence of parity breaking
terms, this would give an invariant regularization scheme.
Furthermore, we recall that Yang-Mills theory in three dimensions is
a superrenormalizable theory, a property which reduces the number of
divergent integrals. It is thus worth looking at the Feynman
diagrams of the theory. Let us begin with the one loop
ghost-antighost self-energy. It is almost trivial to check that, due
to the transversality of the gluon propagator in the Landau gauge,
the Feynman integral for the ghost self-energy
\begin{equation}
g^{2}\int \frac{d^{3}k}{\left( 2\pi \right) ^{3}}\frac{p_{\mu }(p-k)_{\nu }}{%
\left( p-k\right) ^{2}}\left( \delta _{\mu \nu }-\frac{k_{\mu }k_{\nu }}{%
k^{2}}\right) \frac{1}{k^{2}+m^{2}}\;,  \label{ghs}
\end{equation}
where $p_{\mu }$ stands for the external momentum, is free from
ultraviolet divergences. As a consequence we have that, at one loop
order in $\MSbar$,
\begin{equation}
Z_{c}=Z_{\overline{c}}=1,\;\;\;\;\textrm{at one loop order.}
\label{1gh}
\end{equation}
Analogously, by simple inspection, it turns out that the one loop
correction to the ghost-antighost-gluon vertex is also finite. The
same feature holds for the one loop Feynman diagrams contributing to
the four gluon vertex, from which it follows that in $\MSbar$
\begin{equation}
Z_{g}^{2}Z_{A}^{2}=1,\;\;\;\textrm{at one loop order.} \label{ga}
\end{equation}
Moreover, from equation (\ref{rgh}), we have
\begin{equation}
Z_{A}=1,\;\;\;\textrm{at one loop order,}  \label{1a}
\end{equation}
so that
\begin{equation}
Z_{g}=1,\;\;\;\textrm{at one loop order}  \label{1g}
\end{equation}
in $\MSbar$. We see therefore that, at one loop order, the theory is
completely free from ultraviolet divergences, a feature which also
holds in the presence of massless fermions. At higher orders,
ultraviolet divergences could show up.

To provide a non-trivial check of the validity of the relation
(\ref{g1}) from another point of view, we shall make use of the
large $N_{\!f}$ expansion, given the existence of a fixed point in
the $\beta$-function. Within this large $N_{\!f}$ expansion
technique, it is commonly known that this fixed point can be
obtained by analytic continuation of the one existing in
$d=4-2\epsilon$ dimensions. This will be considered in the following
section.

\section{Large $N_{\!f}$ verification.}
Having established the renormalizability of the mass operator in the
Landau gauge, we verify the result in QCD using the large $N_{\!f}$
critical
point method developed in \cite{Vasiliev:yc,Vasiliev:dg} for the non-linear $%
\sigma $ model and extended to QED and QCD in
\cite{Gracey:1991xf,Gracey:iu,Gracey:ua,Gracey:1996up}. Briefly,
this method allows one to compute the critical exponents associated
with the renormalization of the fields, coupling constants or
composite operators at the $d$-dimensional fixed point of the QCD
$\beta $-function. The critical exponents encode all orders
information on the respective anomalous dimensions, $\beta
$-function and operator anomalous dimensions and are more
fundamental than their associated renormalization group functions in
that they are renormalization group invariant. Knowing the explicit
location of the $d$-dimensional fixed point allows one to convert
the information encoded in the exponents to the explicit
coefficients in the four dimensional perturbative expansion of the
renormalization group functions. Since we are interested in the
renormalization of $\frac{1}{2}A_\mu^a A_\mu^a$ in the Landau gauge
and its connection with the gluon and ghost wave function
renormalization we will show that, in agreement with eqs.(\ref{rm}),
(\ref{rm1}), the critical exponent associated with the Landau gauge
renormalization of $A_\mu^a A_\mu^a$ at leading order in large
$N_{\!f}$ is simply the sum of the gluon and ghost wave function
critical exponents. The latter have already been determined in
\cite{Gracey:ua}. Moreover, since the computation is in
$d$-dimensions, $2$~$<$~$d$~$<$~$4$, the three dimensional result of
the previous sections will emerge \emph{naturally}.

To fix notation for this section, we recall that the $d$-dimensional $\overline{\mbox{MS}}$ QCD $\beta $-function, \cite{Tarasov:au}%
, is
\begin{eqnarray}\label{beta}
\beta (a) &=&(d-4)a+\left[
\frac{2}{3}T_{F}N_{\!f}-\frac{11}{6}C_{A}\right]
a^{2}  +\left[ \frac{1}{2}C_{F}T_{F}N_{\!f}+\frac{5}{6}C_{A}T_{F}N_{\!f}-\frac{17%
}{12}C_{A}^{2}\right] a^{3}  \nonumber \\
&&-~\left[ \frac{11}{72}C_{F}T_{F}^{2}N_{\!f}^{2}+\frac{79}{432}%
C_{A}T_{F}^{2}N_{\!f}^{2}+\frac{1}{16}C_{F}^{2}T_{F}N_{\!f}+\frac{2857}{1728}%
C_{A}^{3}\right.   \nonumber \\
&&\left. ~~~~~-~ \frac{205}{288}C_{F}C_{A}T_{F}N_{\!f}-\frac{1415}{864}%
C_{A}^{2}T_{F}N_{\!f}\right] a^{4}+O(a^{5})\;,
\end{eqnarray}
where the group Casimirs are defined by $T^{a}T^{a}$ $=$ $C_{F}I$, $%
f^{acd}f^{bcd}$~$=$~$C_{A}\delta ^{ab}$ and  $\mbox{Tr}\left(
T^{a}T^{b}\right) $~$=$~$T_{F}\delta ^{ab}$. The leading $O(a)$ term
corresponds to the dimension of the coupling in $d$-dimensions and
is necessary to deduce the location of the non-trivial
$d$-dimensional fixed point $a_{c}$. Expanding in powers of
$1/N_{\!f}$ it is given by
\begin{eqnarray}\label{fix}
a_{c} &=&\frac{3\epsilon }{T_{F}N_{\!f}}+\frac{1}{4T_{F}^{2}N_{\!f}^{2}}%
\left[ \frac {} {}33C_{A}\epsilon -\left( 27C_{F}+45C_{A}\right)
\epsilon
^{2}\right.   \nonumber \\
&&+~\left. \left( \frac{99}{4}C_{F}+\frac{237}{8}C_{A}\right)
\epsilon ^{3}+O(\epsilon ^{4})\right] +O\left(
\frac{1}{N_{\!f}^{3}}\right)\;,
\end{eqnarray}
where $d=4-2\epsilon$. QCD is in the same universality class as the
non-Abelian Thirring model (NATM), \cite{Hasenfratz:1992jv}, which
has the Lagrangian
\begin{equation}
\mathcal{L}^{\mbox{\footnotesize{NATM}}}~=~i\bar{\psi}^{i}\partial \!\!\!/\psi ^{i}~+~%
\frac{\lambda ^{2}}{2}\left( \bar{\psi}^{i}\gamma ^{\mu }T^{a}\psi
^{i}\right) ^{2}\;,
\end{equation}
or rewriting it in terms of an auxiliary vector field,
$\tilde{A}_{\mu }^{a}$,
\begin{equation}
\mathcal{L}^{\mbox{\footnotesize{NATM}}}~=~i\bar{\psi}^{i}\partial \!\!\!/\psi ^{i}~+~%
\tilde{A}_{\mu }^{a}\bar{\psi}^{i}\gamma ^{\mu }T^{a}\psi ^{i}~-~\frac{(%
\tilde{A}_{\mu }^{a})^{2}}{2\lambda ^{2}} \;, \label{natmlag}
\end{equation}
where the coupling constant $\lambda $ is dimensionless in two
dimensions. By analogy the NATM plays the same role as the $O(N)$
nonlinear $\sigma $
model in the $d$-dimensional critical point equivalence with the $4$%
-dimensional $O(N)$ $\phi ^{4}$ theory at the $d$-dimensional
Wilson-Fisher fixed point. One feature of the universality criterion
at criticality is that the interactions of the fields play the major
role. Hence, comparing the QCD and NATM Lagrangians where for this
section we take
\begin{equation}
\mathcal{L}^{\mbox{\footnotesize{QCD}}}~=~i\bar{\psi}^{i}\partial \!\!\!/\psi ^{i}~+~%
\tilde{A}_{\mu }^{a}\bar{\psi}^{i}\gamma ^{\mu }T^{a}\psi ^{i}~-~\frac{%
(F_{\mu \nu }^{a})^{2}}{4g^{2}}\;,
\end{equation}
the quark-gluon $3$-point interaction of both models is dominant in
the large $N_{\!f}$ critical point method. In QCD the field strength
of the Lagrangian is infrared irrelevant and drops out of the large
$N_{\!f}$ analysis. However, in practice the triple and quartic
gluon interactions emerge in diagrams with closed quark loops with
respectively three and four external $\tilde{A}_{\mu }^{a}$ fields,
\cite{Hasenfratz:1992jv,Gracey:1996up}. It is worth noting that in
this section alone we have redefined the gluon field and
incorporated a power of the QCD coupling constant into its
definition, $\tilde{A}_{\mu }^{a}$~$=$~$gA_{\mu }^{a}$ which is the
origin of the power of $g^{2}$ factor with the field strength term.
This rescaling is necessary for the application of the critical
point large $N_{\!f}$ programme which requires a unit coupling
constant for the quark gluon interaction and therefore defines the
canonical scaling dimensions in such a way as to make
the calculational tool of uniqueness applicable which was used extensively in the original large $%
N_{\!f}$ critical point method of \cite{Vasiliev:yc,Vasiliev:dg}.
\begin{figure}[ht]
\centering \epsfig{file=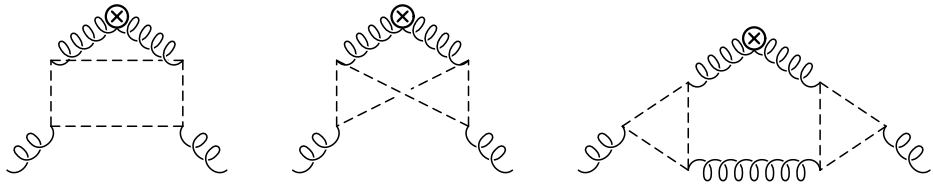,width=12cm}
\caption{$O(1/N_{\!f})$ diagrams contributing to $\eta_{A^2}$.}
\end{figure}
As we are interested in the critical exponents and therefore the
anomalous dimensions of the composite operator $\frac{1}{2}A_\mu^a
A_\mu^a$ in the large $N_{\!f}$ expansion, we follow the method of
\cite{Gracey:1996up}. There the critical exponent $\omega$
associated with the QCD $\beta$-function was computed at
$O(1/N_{\!f})$ in $d$-dimensions by inserting the composite operator
$\left( F^a_{\mu\nu} \right)^2$ into a gluon $2$-point function and
applying the method of \cite{Vasiliev:mq} to determine the critical
dimension of its associated coupling. For the anomalous dimension of
$\frac{1}{2}A_\mu^a A_\mu^a$ we
follow the same approach and note that the appropriate $O(1/N_{\!f})$ large $%
N_{\!f}$ diagrams are given in figure 1 where a gluon line counts
one power of $1/N_{\!f}$ which is why there are two \emph{and} three
loop Feynman diagrams at this order. The latter three loop graph in
fact contains the relevant contribution from the triple gluon vertex
which is absent in the NATM Lagrangian. Unlike perturbation theory
the propagators of figure 1 are not the usual ones. Their asymptotic
scaling forms are deduced from dimensional analysis and consistency
with Lorentz symmetry. In the Landau gauge we have,
\cite{Gracey:iu,Gracey:ua},
\begin{equation}
\psi(k) ~\sim~ \frac{Ak \! \! \! /}{(k^2)^{\mu-\alpha}} ~~,~~
A_{\mu\nu}(k)
~\sim~ \frac{B}{(k^2)^{\mu-\beta}}\left[ \eta_{\mu\nu} - \frac{k_\mu k_\nu}{%
k^2} \right] ~~,~~ c(k) ~\sim~ \frac{C}{(k^2)^{\mu-\gamma}}\;,
\label{nfprops}
\end{equation}
in momentum space at leading order as $k^2\rightarrow\infty$ as one
approaches the $d$-dimensional fixed point. We have given the ghost
propagator asymptotic scaling form for completeness and to define
its scaling dimension even though it is not needed at $O(1/N_{\!f})$
for the explicit computation of the critical exponent of
$\frac{1}{2}A_\mu^aA_\mu^a$. The powers of the propagators are
defined as
\begin{equation}
\alpha ~=~ \mu ~+~ \mbox{\small{$\frac{1}{2}$}} \eta ~~,~~ \beta ~=~
1 ~-~ \eta ~-~ \chi ~~,~~ \gamma ~=~ \mu - 1 +
\mbox{\small{$\frac{1}{2}$}} \eta_c\;, \label{expdefn}
\end{equation}
where $A$, $B$ and $C$ are the momentum independent amplitudes
though only the combinations $z=A^2B$ and $y=C^2B$ appear in
calculations, \cite{Gracey:ua}. We use $\mu=d/2$ for shorthand,
$\eta$ is the critical exponent of the quark field, $\chi$ is the
critical exponent of the quark-gluon vertex anomalous dimension and
$\eta_c$ is the ghost critical exponent. We note that the explicit
$O(1/N_{\!f})$ values of the critical exponents in $d$-dimensions in
the Landau gauge are, \cite{Gracey:ua},
\begin{eqnarray}
\eta_1 &=& \frac{(2\mu-1)(\mu-2)\Gamma(2\mu) C_F}{4\Gamma^2(\mu)
\Gamma(\mu+1) \Gamma(2-\mu) T_F} ~\equiv~ \eta^{\mbox{o}}_1
\frac{C_F}{T_F}\;,
\nonumber \\
\chi_1 &=& - \, \left[ C_F + \frac{C_A}{2(\mu-2)} \right] \frac{\eta^{%
\mbox{o}}_1}{T_F} ~~,~~ \eta_{c \, 1} ~=~ - \, \frac{C_A\eta^{\mbox{o}}_1}{%
2(\mu-2)T_F}  \;,\label{expvals}
\end{eqnarray}
where we will use the notation $\eta=\sum_{i=1}^\infty
\eta_i/N_{\!f}^i $. The expression for the ghost anomalous dimension
follows from the usual Slavnov-Taylor identity as expressed in
exponent language,
\begin{equation}
\eta_c ~=~ \eta ~+~ \chi ~-~ \chi_c\;,
\end{equation}
where $\chi_c$ is the anomalous dimension of the ghost-gluon vertex
and was shown in \cite{Gracey:ua} to vanish in the Landau gauge at
$O(1/N_{\!f})$.

The explicit computation of the exponent associated with the
renormalization
of $\frac{1}{2}A_\mu^a A_\mu^a$, which we will call $%
\eta_{A^2}$, is deduced by inserting (\ref{nfprops}) into the
diagrams of figure 1 and applying the procedure of
\cite{Vasiliev:mq} to determine the scaling dimension of the
operator insertion, $\eta_{\mathcal{O}}$. The value of $\eta_{A^2}$
is deduced from the relation
\begin{equation}
\eta_{A^2} ~=~ \eta ~+~ \chi ~+~ \eta_{\mathcal{O}}\;,
\label{scalreln}
\end{equation}
where the first two terms correspond to the anomalous part of the
gluon critical dimension or wave function renormalization. For
completeness we note that the corresponding critical exponent in the
Thirring model, $\omega^{\mbox{\footnotesize{NATM}}}$, is deduced by
dimensionally analysing the final term of (\ref{natmlag}) giving
\begin{equation}
\omega^{\mbox{\footnotesize{NATM}}} ~=~ \mu ~-~ 1 ~+~ \eta ~+~ \chi
~+~ \eta_{\mathcal{O}} ~.
\end{equation}
In practice a regularization has to be introduced for the Feynman
integrals
which is obtained by shifting the exponent of the vertex renormalization, $%
\chi$, to the new value of $\chi+\Delta$. Here $\Delta$ plays a role
akin to $\epsilon$ in dimensional regularization. Though it should
be stressed that we are working in fixed dimensions, $d$, and not
dimensionally regularizing here. The actual contribution to
$\eta_{\mathcal{O}}$ is determined from the residue of the simple
pole in $\Delta$ from the sum of all the diagrams of figure 1. In
\cite{Gracey:1996up} the two and three loop diagrams were computed
using various techniques such as integration by parts and
uniqueness, \cite{Vasiliev:dg}, after the regularized Feynman
integrals were broken up into a set of basic integrals which were
straightforward to determine and a set which required a substantial
amount of effort particularly in the case of the three loop diagram.
We have used the same integrals here but supplemented with an extra
set since the operator insertion of $\frac{1}{2}A_\mu^a A_\mu^a$
alters the power of the internal gluon line containing the operator
insertion.
\begin{figure}[ht]
\centering \epsfig{file=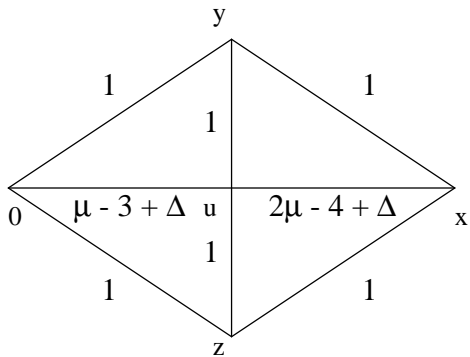,height=5cm}
\caption{Basic three loop Feynman diagram.}
\end{figure}
An example of one of the tedious graphs in this respect is that
illustrated in figure 2 where we have indicated the power of the
propagator beside the line. We have used coordinate space
representation where one integrates over the location of the
internal vertices, $u$, $y$ and $z$, but with $x$ corresponding to
the external coordinate or momentum of the diagram. To determine the
residue with respect to the $\Delta$-pole we convert the integral to
momentum representation, \cite{Vasiliev:dg}, which produces the
first diagram of figure 3. There we have nullified the
regularization since the associated factor from the transformation
is
\begin{equation}
\frac{a^6(1) a(\mu-3) a(2\mu-4)}{a(-1+2\Delta)}\;,
\end{equation}
which, due to the denominator factor, is clearly divergent as $\Delta$~$%
\rightarrow$~$0$ since
$a(\alpha)$~$=$~$\Gamma(\mu-\alpha)/\Gamma(\alpha)$. To proceed we
use the language of \cite{Vasiliev:dg} and apply a conformal
transformation to the first diagram of figure 3 based on the left
external point. Then integrating the unique triangle and subsequent
unique vertex before undoing the original conformal transformation
finally produces the second diagram of figure $3$. The factor
associated with these manipulations from the first diagram of figure
3 is $a^4(\mu-1)/a(2\mu-4)$. To deduce the value of the final
diagram which is $\Delta$-finite we integrate by parts on the top
right internal vertex based on the line with exponent 1. This
produces four two loop diagrams. However, these intermediate
diagrams are in fact divergent though their sum is finite. To ensure
the correct finite part emerges, one introduces a temporary
intermediate regularization prior to integrating by parts by
shifting the exponent of the line labeled $3-\mu$ to an exponent of
$3-\mu+\delta$. In fact two of the resulting diagrams then cancel
exactly, leaving two integrals which are related to the function
$\mbox{ChT}(\alpha,\beta)$ defined in \cite{Vasiliev:dg} and
evaluated exactly in \cite{Chetyrkin:pr,Vasiliev:dg}. Explicitly one
has the
difference of $\mbox{ChT}(-1-\delta,3-\mu)$ and $\mbox{ChT}%
(\mu-3-\delta,3-\mu)$ and expanding in powers of $\delta$ a finite
expression emerges. Accumulating all the contributions the final
contribution of the integral of figure 2 to the critical exponent
computation is
\begin{equation}
\frac{(2\mu-3)(\mu-1)^2(2\mu^2-7\mu+4)}{4\Gamma(\mu+1)\Delta}\;.
\end{equation}

\begin{figure}[th]
\centering \epsfig{file=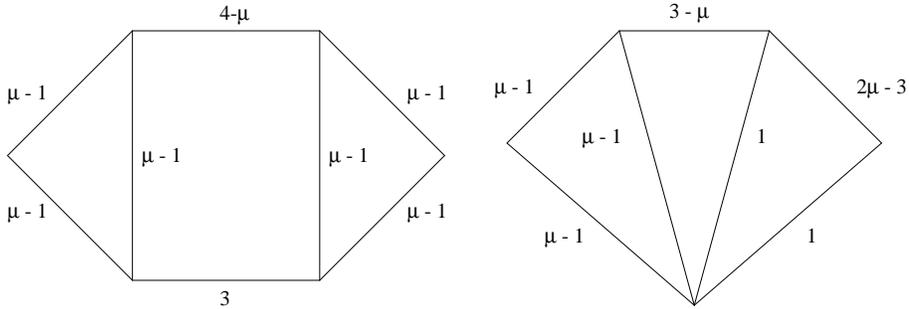,width=12.5cm}
\caption{Intermediate three loop Feynman diagrams.}
\end{figure}
Having completed the computation of all the intermediate basic
integrals we note that the transverse contribution of each of the
four diagrams of figure 1 to $\eta _{\mathcal{O}}$ are respectively,
\begin{eqnarray}
&&-~\frac{(2\mu -1)(2\mu -3)C_{F}\eta _{1}^{\mbox{o}}}{2(\mu -2)T_{F}}~~,~~%
\frac{(2\mu -1)(2\mu -3)[C_{F}-\mbox{\small{$\frac{1}{2}$}}C_{A}]\eta _{1}^{%
\mbox{o}}}{2(\mu -2)T_{F}}~~,  \nonumber \\
&&\frac{(\mu -1)^{2}C_{A}\eta _{1}^{\mbox{o}}}{(\mu -2)T_{F}}\;.
\end{eqnarray}
Hence,
\begin{equation}
\eta _{A^{2}}~=~-~\frac{C_{A}\eta _{1}^{\mbox{o}}}{4(\mu -2)T_{F}N_{\!f}}%
~+~O\left( \frac{1}{N_{\!f}^{2}}\right)\;,
\end{equation}
in $d$-dimensions. Clearly this is equivalent to the sum of
anomalous
dimension parts of the Landau gauge gluon and ghost critical exponents at $%
O(1/N_{\!f})$. More explicitly, from (\ref{expdefn}) and
(\ref{expvals}),
\begin{equation}
\eta _{A^{2}\,1}~=~\eta _{1}~+~\chi
_{1}~-~\mbox{\small{$\frac{1}{2}$}}\eta _{c\,1} \;,\label{stnfreln}
\end{equation}
which due to our choice of conventions and notation was the way this
identity was originally uncovered in \cite{Gracey:2002yt} prior to
the all orders proof of \cite{Dudal:2002pq} and its subsequent
expression in the form of (\ref{g1}). Therefore, (\ref{stnfreln}) is
an explicit $d$-dimensional verification of the all orders result of
the previous section. Moreover, it nicely recovers the
$d$-dimensional case of
\cite{Gracey:2002yt,Dudal:2002pq,Chetyrkin:2004mf}.

As three dimensional QCD is of interest in other problems, we note
that the explicit three dimensional value of
$\omega^{\mbox{\footnotesize{NATM}}}$ is
\begin{equation}
\left. \frac{}{} \omega^{\mbox{\footnotesize{NATM}}} \right|_{d=3} ~=~ \frac{%
1}{2} ~-~ \frac{4C_A}{3\pi^2T_FN_{\!f}} ~+~ O \left(
\frac{1}{N_{\!f}^2} \right)\;.
\end{equation}
In two dimensions, interestingly the critical exponent does not run
to its mean field value and one has
\begin{equation}
\left. \frac{}{} \omega^{\mbox{\footnotesize{NATM}}} \right|_{d=2}
~=~ -~ \frac{C_A}{16T_FN_{\!f}} ~+~ O \left( \frac{1}{N_{\!f}^2}
\right) \;.
\end{equation}

\section{Generalization to other gauges: the example of the Curci-Ferrari
gauge.} The mass operator $A_{\mu }^{a}A_{\mu }^{a}$ in the Landau
gauge can be generalized to other gauges, such as the Curci-Ferrari
and the maximal Abelian gauge. In this case the mixed gluon-ghost
mass operator $\left( \frac{1}{2}A_{\mu }^{a}A_{\mu }^{a}+\alpha
\overline{c}^{a}c^{a}\right) $ has to be considered, where $\alpha $
stands for the gauge parameter. Let us consider here the case of the
Curci-Ferrari nonlinear gauge. For the gauge fixed action we have
\begin{eqnarray}
S_{\mathrm{CF}} &=&\int d^{3}x\,\left( -~\frac{1}{4}F_{\mu \nu
}^{a}F_{\mu
\nu }^{a}+b^{a}\partial _{\mu }A_{\mu }^{a}+\,\frac{\alpha }{2}b^{a}b^{a}+%
\overline{c}^{a}\partial _{\mu }\left( D_{\mu }c\right) ^{a}-\frac{\alpha }{2%
}gf^{abc}b^{a}\overline{c}^{b}c^{c}\right.   \nonumber \\ &-&\left.\frac{\alpha }{8}g^{2}f^{abc}f^{cde}\overline{c}^{a}%
\overline{c}^{b}c^{d}c^{e}+m^{2}\left( \frac{1}{2}A_{\mu }^{a}A_{\mu
}^{a}+\alpha \overline{c}^{a}c^{a}\right) \right) \;.  \label{cf}
\end{eqnarray}
Notice that in this case also the Faddeev-Popov ghosts $\overline{c}^{a}$,$%
\;c^{a}$ are massive. Moreover, the Curci-Ferrari gauge reduces to
the Landau gauge in the limit $\alpha \rightarrow 0$. The action
(\ref{cf})\ is
invariant under the BRST and $\delta $ transformations of eqs.(\ref{sm}), (%
\ref{d}). Introducing the external action
\begin{equation}
S_{\mathrm{ext}}=\int d^{3}x\left( -~\Omega _{\mu }^{a}\,\left(
D_{\mu }c\right) ^{a}+\,L^{a}\,\frac{g}{2}f^{abc}c^{b}c^{c}\right)
\;,  \nonumber
\end{equation}
it follows that the complete classical action
\begin{equation}
\Sigma _{\mathrm{CF}}~=~S_{\mathrm{CF}}+~S_{\mathrm{ext}}\;,
\label{cacf}
\end{equation}
turns out to be constrained by the Slavnov-Taylor identity
\begin{equation}
\mathcal{S}(\Sigma )=\int d^{3}x\,\left( \frac{\delta \Sigma
}{\delta \Omega
_{\mu }^{a}}\frac{\delta \Sigma }{\delta A_{\mu }^{a}}+\frac{\delta \Sigma }{%
\delta L^{a}}\frac{\delta \Sigma }{\delta c^{a}}+b^{a}\frac{\delta \Sigma }{%
\delta \overline{c}^{a}}-m^{2}c^{a}\frac{\delta \Sigma }{\delta b^{a}}%
\right) =0\;,  \label{stcf}
\end{equation}
and by the $\delta $ Ward identity
\begin{equation}
\mathcal{W}(\Sigma )=\int d^{3}x\,\left( c^{a}\frac{\delta \Sigma
}{\delta
\overline{c}^{a}}+\frac{\delta \Sigma }{\delta L^{a}}\frac{\delta \Sigma }{%
\delta b^{a}}\right) \;=0\;.  \label{dwcf}
\end{equation}
Due to the presence of the quartic ghost-antighost term  $g^{2}f^{abc}f^{cde}%
\overline{c}^{a}\overline{c}^{b}c^{d}c^{e}$ and of $gf^{abc}b^{a}\overline{c}%
^{b}c^{c}$ the additional Ward identities (\ref{beq}) and (\ref{gh})
of the Landau gauge do not hold in the present case. Nevertheless,
identities (\ref{stcf}) and (\ref{dwcf}) ensure the multiplicative
renormalizability of the model. Proceeding as in the previous
section, it turns out that the most
general invariant counterterm contains five free independent parameters, $%
\sigma $, $a_{1}$, $a_{2}$, $a_{3}$, $a_{5}$ and is given by
\begin{eqnarray}
\Sigma _{\mathrm{CF}}^{\mathrm{count}} &=&\int d^{3}x\left(
-\frac{(\sigma +4a_{1})}{4}F_{\mu \nu }^{a\,}F_{\mu \nu
}^{a}+a_{1}F_{\mu \nu }^{a\,}\partial _{\mu }A_{\nu
}^{a}+(a_{1}-a_{2})b^{a}\partial _{\mu }A_{\mu }^{a}\newline
\right.   \nonumber \\
&+&a_{1}(\partial _{\mu }\bar{c}^{a}+\Omega _{\mu }^{a})\partial
_{\mu }c^{a}+(a_{1}-a_{2})\bar{c}^{a}\partial _{\mu }(D_{\mu
}c)^{a}+a_{5}(\partial _{\mu }\bar{c}^{a}+\Omega _{\mu }^{a})(D_{\mu }c)^{a}%
\nonumber \\
&-&a_{3}\frac{\alpha }{2}b^{a}b^{a}+\frac{(a_{3}+a_{5})}{2}\alpha
gf^{abc}b^{a}\bar{c}^{b}c^{c}+\frac{(a_{3}+2a_{5})}{8}\alpha
g^{2}f^{abc}f^{cde}\bar{c}^{a}\bar{c}^{b}c^{d}c^{e}
\nonumber \\
&-&\left. \frac{a_{5}}{2}gf^{abc}L^{a}c^{b}c^{c}+m^{2}\left( (a_{1}-\frac{%
a_{2}}{2}+\frac{a_{5}}{2})A_{\mu }^{a}A_{\mu }^{a}-\alpha a_{3}\bar{c}%
^{a}c^{a}\right) \right) \;.   \label{ctcf}
\end{eqnarray}
The parameters $\sigma $, $a_{1}$, $a_{2}$, $a_{3}$, $a_{5}$ are
easily seen to correspond to a multiplicative renormalization of the
fields, sources and parameters, according to
\begin{eqnarray}
Z_{g} &=&1-\eta \frac{\sigma }{2}\,\,\,\,,  \nonumber \\
Z_{A}^{1/2} &=&1+\eta \left( \frac{\sigma }{2}+a_{1}\right) \;,
\nonumber \\
Z_{c}^{1/2} &=&Z_{\overline{c}}^{1/2}=1-\eta \left( \frac{\left(
a_{2}+a_{5}\right) }{2}\right) \,\,,  \nonumber \\
Z_{L} &=&1+\eta \left( \frac{\sigma }{2}+a_{2}\right) \,\,,
\nonumber
\\
Z_{\alpha } &=&1+\eta \left( -a_{3}+2a_{2}+\sigma \right) \;,
\label{zcf}
\end{eqnarray}
and
\begin{eqnarray}
Z_{\Omega } &=&{Z}_{A}{^{-1/2}Z}_{c}^{1/2}Z_{L}{\;,\ \newline
}  \nonumber \\
Z_{b}^{1/2} &=&Z_{L}^{-1}\;,  \nonumber \\
Z_{m^{2}} &=&Z_{L}^{-2}Z_{c}^{-1}\;.  \label{zmcf}
\end{eqnarray}
In particular, from eqs.(\ref{zmcf}) it follows that the
renormalization factor $Z_{m^{2}}$ is not independent, being
expressed in terms of the
ghost renormalization factor $Z_{c}$\ and of the renormalization factor $%
Z_{L}$ of the source $L^{a}$ coupled to the composite ghost operator $\frac{%
1}{2}gf^{abc}c^{b}c^{c}$. Again, these results are in complete
agreement with those obtained in the four dimensional case
\cite{Dudal:2003pe}.

\section{Conclusion.}
In this paper we have analysed the renormalization properties of the
mass operator $A_{\mu }^{a}A_{\mu }^{a}$ in three dimensional
Yang-Mills theories in the Landau gauge. In analogy with the four
dimensional case, the renormalization factor $Z_{m^{2}}$ is not an
independent parameter of the theory, as expressed by the relations
(\ref{rm}) and (\ref{rm1}), which have been explicitly verified in
the large $N_{\!f}$ expansion method. These results will be used in
order to investigate by analytical methods the possible formation of
the gauge condensate $\left\langle A_{\mu }^{a}A_{\mu
}^{a}\right\rangle $. This would provide a dynamical generation of a
parity preserving mass for the gluons in three dimensions, a topic
which has been extensively investigated in recent years. For
instance, see
\cite{Karabali:1995ps,Karabali:1996je,Jackiw:1995nf,Jackiw:1997jg}.

Finally, we underline that the Curci-Ferrari gauge allows one to
study the generalized mixed gluon-ghost condensate $\left\langle
\frac{1}{2}A_{\mu }^{a}A_{\mu }^{a}+\alpha
\overline{c}^{a}c^{a}\right\rangle $. In particular, as discussed in
the four dimensional case, the presence of the gauge parameter
$\alpha $ could be useful to investigate the gauge independence of
the vacuum energy, due to the formation of the aforementioned
condensates.

\section*{Acknowledgments.}
D.~Dudal and S.~P.~Sorella are grateful to D.~Anselmi for useful
discussions. The Conselho Nacional de Desenvolvimento Cient\'{i}fico
e Tecnol\'{o}gico
(CNPq-Brazil), the SR2-UERJ and the Coordena{\c{c}}{\~{a}}o de Aperfei{\c{c}}%
oamento de Pessoal de N{\'\i}vel Superior (CAPES) are gratefully
acknowledged for financial support. D.~Dudal would like to
acknowledge the warm hospitality at the Physics Institute of the
UERJ, where part of this work was done. R.~F.~Sobreiro would like to
thank the Department of Mathematical Physics and Astronomy of the
Ghent University, where part of this work was completed.


\begin{thebibliography}{99}
\bibitem{Verschelde:2001ia}
H.~Verschelde, K.~Knecht, K.~Van Acoleyen and M.~Vanderkelen, Phys.\
Lett.\ B \textbf{516} (2001) 307.

\bibitem{Browne:2003uv}
R.~E.~Browne and J.~A.~Gracey, JHEP \textbf{0311} (2003) 029.

\bibitem{Langfeld:2001cz}
K.~Langfeld, H.~Reinhardt and J.~Gattnar, Nucl.\ Phys.\ B
\textbf{621} (2002) 131.

\bibitem{Aguilar:2004kt}
A.~C.~Aguilar and A.~A.~Natale, hep-ph/0405024.

\bibitem{Dudal:2002pq}
D.~Dudal, H.~Verschelde and S.~P.~Sorella, Phys.\ Lett.\ B
\textbf{555} (2003) 126.

\bibitem{Gracey:2002yt}
J.~A.~Gracey, Phys.\ Lett.\ B \textbf{552} (2003) 101.

\bibitem{Dudal:2003np}
D.~Dudal, H.~Verschelde, V.~E.~R.~Lemes, M.~S.~Sarandy,
R.~F.~Sobreiro, S.~P.~Sorella and J.~A.~Gracey, Phys.\ Lett.\ B
\textbf{574} (2003) 325.

\bibitem{Dudal:2003by}
D.~Dudal, H.~Verschelde, J.~A.~Gracey, V.~E.~R.~Lemes,
M.~S.~Sarandy, R.~F.~Sobreiro and S.~P.~Sorella, JHEP \textbf{0401}
(2004) 044.

\bibitem{Kondo:2001nq}  K.~I.~Kondo,
Phys.\ Lett.\ B \textbf{514} (2001) 335.

\bibitem{Kondo:2001tm}
K.~I.~Kondo, T.~Murakami, T.~Shinohara and T.~Imai, Phys.\ Rev.\ D
\textbf{65} (2002) 085034.

\bibitem{Dudal:2003pe}
D.~Dudal, H.~Verschelde, V.~E.~R.~Lemes, M.~S.~Sarandy,
R.~F.~Sobreiro, S.~P.~Sorella, M.~Picariello, J.~A.~Gracey, Phys.\
Lett.\ B \textbf{569} (2003) 57.

\bibitem{Dudal:2003gu}
D.~Dudal, H.~Verschelde, V.~E.~R.~Lemes, M.~S.~Sarandy,
S.~P.~Sorella and M.~Picariello, Annals Phys.\ \textbf{308} (2003)
62.

\bibitem{Dudal:2004rx}
D.~Dudal, J.~A.~Gracey, V.~E.~R.~Lemes, M.~S.~Sarandy,
R.~F.~Sobreiro, S.~P.~Sorella and H.~Verschelde, hep-th/0406132.

\bibitem{Appelquist:1981vg}
T.~Appelquist and R.~D.~Pisarski, Phys.\ Rev.\ D \textbf{23} (1981)
2305.

\bibitem{Jackiw:1980kv}
R.~Jackiw and S.~Templeton, Phys.\ Rev.\ D \textbf{23} (1981) 2291.

\bibitem{Deser:1981wh}
S.~Deser, R.~Jackiw and S.~Templeton, Annals Phys.\  {\bf 140}
(1982) 372 [Erratum-ibid.\  {\bf 185} (1988\ APNYA,281,409-449.2000)
406.1988\ APNYA,281,409].

\bibitem{Delduc:1989uc}
F.~Delduc and S.~P.~Sorella, Phys.\ Lett.\ B \textbf{231} (1989)
408.

\bibitem{Dudal:2002ye}
D.~Dudal, V.~E.~R.~Lemes, M.~S.~Sarandy, S.~P.~Sorella and
M.~Picariello, JHEP \textbf{0212} (2002) 008.

\bibitem{Piguet:1995er}
O.~Piguet and S.~P.~Sorella, Lect.\ Notes Phys.\ \textbf{M28} (1995)
1.

\bibitem{Blasi:1990xz}
A.~Blasi, O.~Piguet and S.~P.~Sorella, Nucl.\ Phys.\ B \textbf{356}
(1991) 154.

\bibitem{Pisarski:1985yj}
R.~D.~Pisarski and S.~Rao, Phys.\ Rev.\ D {\bf 32} (1985) 2081.

\bibitem{Vasiliev:yc}
A.~N.~Vasiliev, Y.~M.~Pismak and Y.~R.~Honkonen, Theor.\ Math.\
Phys.\ \textbf{46} (1981) 104 [Teor.\ Mat.\ Fiz.\ \textbf{46} (1981)
157].

\bibitem{Vasiliev:dg}
A.~N.~Vasiliev, Y.~M.~Pismak and Y.~R.~Honkonen,
Theor.\ Math.\ Phys.\ \textbf{47} (1981) 465 [Teor.\ Mat.\ Fiz.\ \textbf{47}%
 (1981) 291].

\bibitem{Gracey:1991xf}  J.~A.~Gracey,
J.\ Phys.\ A \textbf{24} (1991) L431.

\bibitem{Gracey:iu}
J.~A.~Gracey, Nucl.\ Phys.\ B \textbf{414} (1994) 614.

\bibitem{Gracey:ua}
J.~A.~Gracey, Phys.\ Lett.\ B \textbf{318} (1993) 177.

\bibitem{Gracey:1996up}
J.~A.~Gracey, Phys.\ Lett.\ B \textbf{373} (1996) 178.

\bibitem{Tarasov:au}
O.~V.~Tarasov, A.~A.~Vladimirov and A.~Y.~Zharkov, Phys.\ Lett.\ B
\textbf{93} (1980) 429.

\bibitem{Hasenfratz:1992jv}
A.~Hasenfratz and P.~Hasenfratz, Phys.\ Lett.\ B \textbf{297} (1992)
166.

\bibitem{Vasiliev:mq}
A.~N.~Vasiliev and M.~Y.~Nalimov, Theor.\ Math.\ Phys.\ \textbf{55}
(1983) 423 [Teor.\ Mat.\ Fiz.\ \textbf{55} (1983) 163].

\bibitem{Chetyrkin:pr}
K.~G.~Chetyrkin, A.~L.~Kataev and F.~V.~Tkachov, Nucl.\ Phys.\ B
\textbf{174} (1980) 345.

\bibitem{Chetyrkin:2004mf}
K.~G.~Chetyrkin, hep-ph/0405193.

\bibitem{Karabali:1995ps}
D.~Karabali and V.~P.~Nair, Nucl.\ Phys.\ B \textbf{464} (1996) 135.

\bibitem{Karabali:1996je}
D.~Karabali and V.~P.~Nair, Phys.\ Lett.\ B \textbf{379} (1996) 141.

\bibitem{Jackiw:1995nf}
R.~Jackiw and S.~Y.~Pi, Phys.\ Lett.\ B \textbf{368} (1996) 131.

\bibitem{Jackiw:1997jg}
R.~Jackiw and S.~Y.~Pi, Phys.\ Lett.\ B {\bf 403} (1997) 297.
\end{thebibliography}
\end{document}